\documentclass{webofc}
\usepackage[varg]{txfonts}   
\usepackage{pythonhighlight}
\usepackage{hyperref}
\begin{document}
\title{The Scikit-HEP Project}
%
%

\author{\firstname{Eduardo} \lastname{Rodrigues}\inst{1}\fnsep\thanks{\email{eduardo.rodrigues@uc.edu}}
}

\institute{University of Cincinnati
          }

\abstract{%
The Scikit-HEP project is a community-driven and community-oriented effort with the aim
of providing Particle Physics at large with a Python scientific toolset containing core and common tools.
The project builds on five pillars that embrace the major topics involved in a physicist’s analysis work:
datasets, data aggregations, modelling, simulation and visualisation.
The vision is to build a user and developer community engaging collaboration across experiments,
to emulate scikit-learn's unified interface with Astropy's embrace of third-party packages,
and to improve discoverability of relevant tools.
}
\maketitle
\section{Introduction}
\label{intro}
\begin{figure}[h]
\centering
\includegraphics[width=7cm,clip]{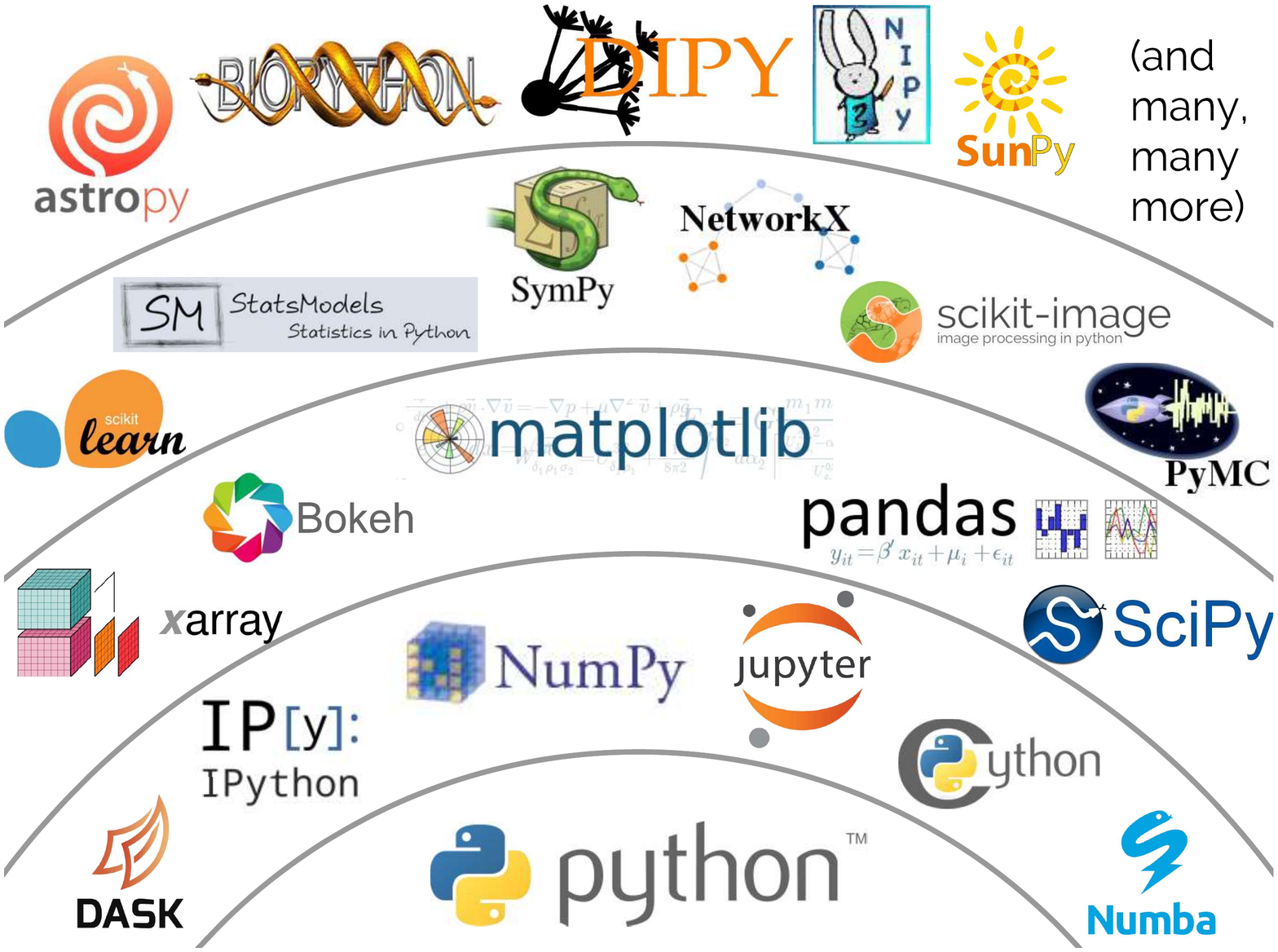}
\caption{Schematic view of the Python scientific software ecosystem.
Figure taken from Jake VanderPlas's presentation at the PyCon 2017 conference~\cite{JakeVanderPlas}.}
\label{fig-JakeVanderPlas}
\end{figure}

It is acknowledged that Python is an extremely popular programming language across a broad range of communities.
Outside High Energy Physics (HEP), the Python scientific ecosystem is built atop the "building blocks"
of the SciPy ecosystem of open-source software for mathematics, science, and engineering~\cite{SciPy}.
A self-explanatory visualisation of the ecosystem is given in figure~\ref{fig-JakeVanderPlas}.
The ecosystem grows to incorporate data manipulation and visualisation tools, packages for statistics and machine learning, etc.
At the top of the "pyramid" lie domain-specific projects -- for example, astropy~\cite{astropy} --
which build on and exploit the building blocks.

Traditionally, HEP has been evolving in a rather disjoint ecosystem based on the C++ ROOT data analysis framework~\cite{ROOT}.
Same as for the Python scientific ecosystem, it provides tools for data manipulation and modeling, for fitting,
for statistics and machine learning applications. But it is a \textit{toolkit} rather than a \textit{toolset},
with bindings to Python.

Various initiatives exist or have existed, which try and link both HEP and non-HEP worlds.
But they mainly tackle(d) specific topics.
We believe there is need for a more generalised effort, domain-specific oriented.

\section{Scikit-HEP project overview}
\label{project}
The Scikit-HEP project~\cite{Scikit-HEP-project} is a community-driven and community-oriented effort with the aim
of providing Particle Physics at large with a Python scientific \textit{toolset} containing core and common tools.
The project builds on five pillars that embrace the major topics involved in a physicist’s analysis work:
datasets, data aggregations, modelling, simulation and visualisation.

The project should neither be seen as a replacement for ROOT nor a replacement for the Python ecosystem based on the SciPy suite.
It is rather the following:
\begin{itemize}
\item An initiative to improve the interoperability between HEP tools and the Python ecosystem,
      expanding the typical set of tools for HEP physicists with common APIs and definitions to ease “cross-talk".
\item An initiative to build a community of developers and users.
\item An effort to improve discoverability of relevant tools.
\end{itemize}

\begin{figure}[h]
\centering
\includegraphics[width=12cm,clip]{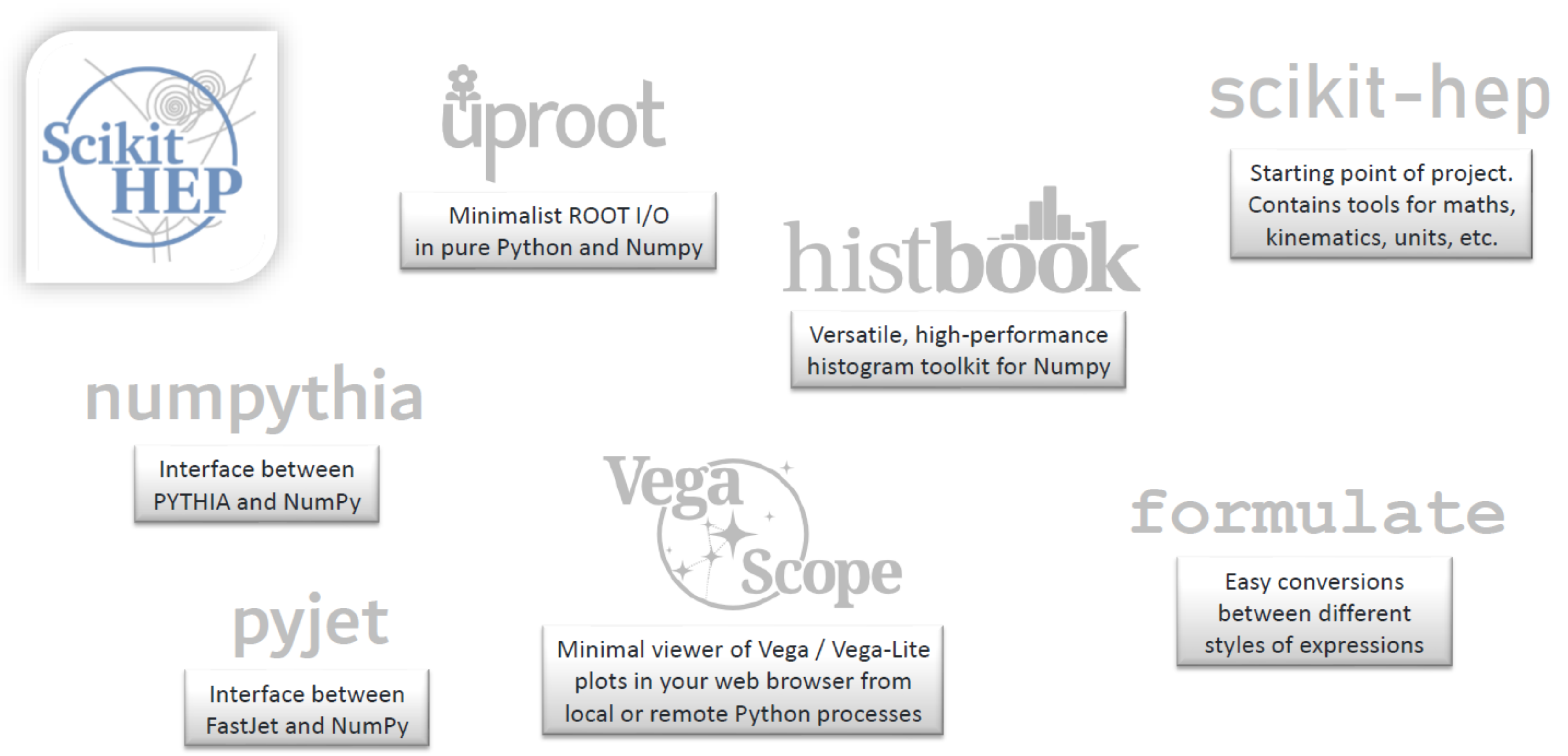}
\caption{Overview of the packages making the Scikit-HEP toolset.
All GitHub repositories can be found at the location \url{https://github.com/scikit-hep}.}
\label{fig-Scikit-HEP}
\end{figure}

The Scikit-HEP toolset is depicted in figure~\ref{fig-Scikit-HEP}.
For completeness, it should be mentioned that the well-known packages \pyth{root_numpy}~\cite{root-numpy}
and \pyth{root_pandas}~\cite{root-pandas}, pre-dating the project, are not described in this report.
They are nevertheless part of the project, but somewhat deprecated by the new and more versatile package
\pyth{uproot}~\cite{uproot}, see below.

The remainder of this report briefly presents each package with simple examples of main functionality.

\section{Scikit-HEP packages walk}
\label{packages}
A ROOT file can natively and trivially be read with the pure I/O \pyth{uproot} package~\cite{uproot}.
The package strength lies in the fact that it does \textit{not} depend on ROOT, making it installable
on virtually any computer via a \pyth{pip install uproot} command.
It bridges the ROOT and the NumPy-powered worlds.
A ROOT \pyth{TTree} object is read as a dictionary of arrays as follows:

\begin{python}
>>> import uproot
>>> rootfile = "Zmumu.root"
>>> zmumu = uproot.open(rootfile)["events"]
>>> zmumu.arrays(["px1","px2","py1","py2","M"])
{b'px1': array([-41.19528764,  35.11804977,  35.11804977, ...,  32.37749196,
         32.37749196,  32.48539387]),
 b'px2': array([ 34.14443725, -41.19528764, -40.88332344, ..., -68.04191497,
        -68.79413604, -68.79413604]),
 b'py1': array([ 17.4332439 , -16.57036233, -16.57036233, ...,   1.19940578,
          1.19940578,   1.2013503 ]),
 b'py2': array([-16.11952457,  17.4332439 ,  17.29929704, ..., -26.10584737,
        -26.39840043, -26.39840043]),
 b'M': array([82.46269156, 83.62620401, 83.30846467, ..., 95.96547966,
        96.49594381, 96.65672765])}
\end{python}

Going beyond simple NumPy arrays, the \texttt{scikit-hep} package~\cite{scikit-hep}
provides dataset classes storing provenance information, where variables are also easily created on the fly:

\begin{python}
>>> from skhep.dataset.numpydataset import *
>>> zmumu_dataset = NumpyDataset(zmumu.arrays(["px1","px2","py1","py2","M"]))
>>> zmumu_dataset
NumpyDataset([(-41.19528764,  34.14443725,  17.4332439 , -16.11952457, 82.46269156),
       ( 35.11804977, -41.19528764, -16.57036233,  17.4332439 , 83.62620401),
       ( 35.11804977, -40.88332344, -16.57036233,  17.29929704, 83.30846467),
       ...,
       ( 32.37749196, -68.04191497,   1.19940578, -26.10584737, 95.96547966),
       ( 32.37749196, -68.79413604,   1.19940578, -26.39840043, 96.49594381),
       ( 32.48539387, -68.79413604,   1.2013503 , -26.39840043, 96.65672765)],
      dtype=[('px1', '<f8'), ('px2', '<f8'), ('py1', '<f8'), ('py2', '<f8'), ('M', '<f8')])
>>>
>>> zmumu_dataset.M
SkhepNumpyArray([82.46269156, 83.62620401, 83.30846467, ..., 95.96547966,
                 96.49594381, 96.65672765])
\end{python}

\begin{python}
>>> zmumu_dataset.pt1 = np.sqrt(zmumu_dataset.px1**2 + zmumu_dataset.py1**2)
>>> print(zmumu_dataset.provenance[1].detail)
Array pt1 has been created (0: <ObjectOrigin>
1: <Transformation(px1 has been squared)>
2: <Transformation(py1**2 has been added to px1**2)>
3: <Transformation(px1**2 + py1**2 has been raised to the power of 0.5)>)
\end{python}

Data selections can be performed in a variety of ways:

\begin{python}
>>> zmumu_dataset.pt2 = np.sqrt(zmumu_dataset.px2**2 + zmumu_dataset.py2**2)
>>> zmumu_dataset1 = zmumu_dataset[(zmumu_dataset.pt1 > 20) & (zmumu_dataset.pt2 > 20)]
>>> # much simpler than line above:
>>> zmumu_dataset2 = zmumu_dataset.select("pt1 > 20 & pt2 > 20")
>>> zmumu_dataset1.nentries
2008
>>> zmumu_dataset2.nentries
2008
\end{python}

The \pyth{formulate}~\cite{formulate} package eases the conversions between different styles
of (selection) expressions:

\begin{python}
>>> import formulate
>>> # write the selection as a formula
>>> pt1 = formulate.from_root('TMath::Sqrt(px1**2 + py1**2)')
>>> pt2 = formulate.from_root('TMath::Sqrt(px2**2 + py2**2)')
>>> cut = (pt1 > 20) & (pt2 > 20)
>>> cut.to_numexpr()
'(sqrt(((px1 ** 2) + (py1 ** 2))) > 20) & (sqrt(((px2 ** 2) + (py2 ** 2))) > 20)'
>>> zmumu_dataset3 = zmumu_dataset.select(cut.to_numexpr())
>>> zmumu_dataset3.nentries
2008
>>> zmumu_dataset3.provenance
0: <ObjectOrigin>
1: <Transformation(Array pt1 has been created)>
2: <Transformation(Array pt2 has been created)>
3: <Transformation(Selection, (sqrt(((px1 ** 2) + (py1 ** 2))) > 20) & (sqrt(((px2 ** 2) + (py2 ** 2))) > 20), applied)>
\end{python}

The manipulation and analysis of datasets typically involves data aggregations and visualisation.
The package \pyth{histbook}~\cite{histbook} provides a versatile, high-performance histogram toolkit for NumPy.
The histograms, often created from ntuples, can be of arbitrary number of dimensions.
The package provides methods for selecting, rebinning, and projecting into lower-dimensional space.
The created histograms are subsequently exportable to a variety of formats, such as ROOT, Pandas, etc.
They can be plotted with Vega-Lite~\cite{VegaLite}, a high-level grammar of interactive graphics
built on top of Vega~\cite{Vega},
a declarative language for creating, saving, and sharing interactive visualisation designs:

\begin{python}
>>> from histbook import *
>>> from vega import VegaLite as canvas
>>> histogram = Hist(bin("Z_M", 50, 0, 125))
>>> M = zmumu_dataset.M.view(np.ndarray)
>>> histogram.fill(Z_M = M)
>>>
>>> histogram.step("Z_M").to(canvas)
>>>
>>> histogram.root()
Welcome to JupyROOT 6.14/00
<ROOT.TH1D object at 0x7f8a305449a0>
>>>
>>> histogram.pandas()
\end{python}

\pyth{VegaScope}~\cite{vegascope} is a minimal viewer of Vega and Vega-Lite graphics from Python.
It provides visualisation in the web browser from local or remote Python processes.

The \pyth{numpythia}~\cite{numpythia} and \pyth{pyjet}~\cite{pyjet} packages provide
interfaces between NumPy and the Pythia~\cite{Pythia} event generator
and the FastJet~\cite{FastJet} jet finding algorithm, respectively.
The collision events generated with Pythia are piped into NumPy arrays:

\begin{python}
>>> from numpythia import Pythia, hepmc_write, hepmc_read
>>> from numpythia import STATUS, HAS_END_VERTEX, ABS_PDG_ID
>>>
>>> params = {"Beams:eCM": 13000, "WeakSingleBoson:ffbar2gmZ": "on", "23:onMode": "off" ,"23:onIfAny": "13", "WeakZ0:gmZmode": 2}
>>>
>>> pythia = Pythia(params=params)
>>> selection = ((STATUS == 1) & ~HAS_END_VERTEX)
>>>
>>> for event in pythia(events=100):
...     array = event.all(selection)
...     muplus = array[array["pdgid"] == 13]  # select muons
\end{python}

\noindent
The \pyth{pyjet} package can trivially receive the inputs from the above
and feed the generated events into FastJet:

\begin{python}
>>> from pyjet import cluster
>>> from pyjet.testdata import get_event
>>>
>>> vectors = get_event()
>>> sequence = cluster(vectors, R=0.1, p=-1)
>>> jets = sequence.inclusive_jets()   # list of PseudoJets
\end{python}

\section{Outlook}
\label{outlook}
The Scikit-HEP project is gaining interest and momentum as a Python library for HEP analysis.
Some of the packages are in fact being used by other communities, in particular the astroparticle physics community.

Much can already be done with the packages described here.
The various packages presented are being further developed and improved as users get to strain test them, and provide feedback.
It is foreseen that other packages join the project in order to complement the offered toolset.

Anyone is welcome to get in touch as user or developer via the public
\href{mailto:scikit-hep-forum@googlegroups.com}{scikit-hep-forum} mailing list.
All project administrators and package maintainers can be reached with the
\href{mailto:scikit-hep-admins@googlegroups.com}{scikit-hep-admins} mailing list.

\section*{Acknowledgements}
This work is supported by the United States National Science Foundation under award number ACI-1450319.
Any opinions, findings, and conclusions or recommendations expressed in this material are those of the author
and do not necessarily reflect the views of the National Science Foundation.

%
%

\end{document}